\documentclass[twocolumn]{aastex631}
\usepackage[T1]{fontenc}
\usepackage{newtxtext,newtxmath}
\usepackage{amsmath}
\usepackage{graphicx}
\usepackage{placeins}
\hypersetup{colorlinks=true,allcolors=blue}

\begin{document}

\title{Resolving the Hubble Tension at $0.1\sigma$ with Early Dark Energy: Key Role of the Highest-redshift Galaxies from JWST Observations}

\author[0009-0005-6921-3201]{Guo-Hong Du}
\affiliation{Liaoning Key Laboratory of Cosmology and Astrophysics, College of Sciences, Northeastern University, Shenyang 110819, China}

\author[0009-0004-6982-4021]{Tian-Nuo Li}\thanks{}
\affiliation{School of Fundamental Physics and Mathematical Sciences, Hangzhou Institute for Advanced Study, UCAS, Hangzhou 310024, China}
\affiliation{Liaoning Key Laboratory of Cosmology and Astrophysics, College of Sciences, Northeastern University, Shenyang 110819, China}

\author[0000-0002-3512-2804]{Jing-Fei Zhang}
\affiliation{Liaoning Key Laboratory of Cosmology and Astrophysics, College of Sciences, Northeastern University, Shenyang 110819, China}

\author[0000-0002-6029-1933]{Xin Zhang}\thanks{}
\affiliation{Liaoning Key Laboratory of Cosmology and Astrophysics, College of Sciences, Northeastern University, Shenyang 110819, China}
\affiliation{MOE Key Laboratory of Data Analytics and Optimization for Smart Industry, Northeastern University, Shenyang 110819, China}
\affiliation{National Frontiers Science Center for Industrial Intelligence and Systems Optimization, Northeastern University, Shenyang 110819, China}

\correspondingauthor{Tian-Nuo Li}
\email{litiannuo@ucas.ac.cn}
\correspondingauthor{Xin Zhang}
\email{zhangxin@mail.neu.edu.cn}

\begin{abstract}
We show that the Hubble tension can be resolved at the $0.1\sigma$ level within the early dark energy (EDE) framework, with the highest-redshift galaxies from the JWST observations playing a key role. By accounting for the reference cosmologies of the JWST ultraviolet luminosity function measurements and combining them with cosmic microwave background (CMB) and DESI baryon acoustic oscillation data, we obtain $H_0=72.76\pm1.00~\mathrm{km\,s^{-1}\,Mpc^{-1}}$ in the axion EDE model, in excellent agreement with SH0ES at the $0.1\sigma$ level. We find that the enhanced galaxy abundances at $z=11$ and $12.5$ play a key role in driving this result. These observations favor stronger small-scale matter power, which can be accommodated by increasing the primordial scalar spectral index $n_{\rm s}$. Under the CMB constraints, the positive correlations among $n_{\rm s}$, $f_{\rm EDE}$, and $H_0$ lead to a higher inferred $H_0$. Allowing additional scale dependence in the primordial power spectrum through $\alpha_{\rm s}$ yields a positive running at more than $4\sigma$, further increasing $H_0$ to $73.75\pm1.04~\mathrm{km\,s^{-1}\,Mpc^{-1}}$, with its value even exceeding the SH0ES measurement. These results reveal how high-redshift small-scale structure observations influence the inferred $H_0$ through the interplay between the primordial power spectrum and EDE parameter degeneracies.
\end{abstract}


\section{Introduction}
\label{Sec1}

The Hubble constant $H_0$ characterizes the present expansion rate of the Universe and is one of the most fundamental parameters in modern cosmology. However, measurements of $H_0$ obtained from different cosmological observations exhibit a significant discrepancy. Within the standard $\Lambda$CDM model, the Planck cosmic microwave background (CMB) observations infer $H_0=67.4\pm0.5~\rm km\,s^{-1}\,Mpc^{-1}$ \citep{Planck:2018vyg}, whereas the SH0ES collaboration obtains $H_0=73.04\pm1.04~\rm km\,s^{-1}\,Mpc^{-1}$ from the local distance ladder \citep{Riess:2021jrx}. The discrepancy reaches more than $5\sigma$ level, constituting the well-known Hubble tension and one of the most prominent challenges to the $\Lambda$CDM model \citep{Zhao:2017urm,Verde:2019ivm,Guo:2018ans,Poulin:2018cxd,Vagnozzi:2019ezj,DiValentino:2020zio,DiValentino:2021izs,Shah:2021onj,Vagnozzi:2021gjh,Gao:2021xnk,Perivolaropoulos:2021jda,Schoneberg:2021qvd,Abdalla:2022yfr,DiValentino:2022fjm,Lee:2022gzh,Kamionkowski:2022pkx,Giare:2023xoc,Hu:2023jqc,Vagnozzi:2023nrq,Bernui:2023byc,Giare:2024akf,Jiang:2025ylr,Pedrotti:2024kpn,Pedrotti:2025ccw,Pedrotti:2026dwj,Du:2026qtq,Cai:2026swf,Schoneberg:2026vaf,Du:2025csv}. This tension has motivated extensive investigations of new physics beyond $\Lambda$CDM, such as early dark energy, interactions within the dark sector, and other extensions \citep{Boisseau:2000pr,Chevallier:2000qy,Li:2004rb,Huang:2004wt,Zhang:2005hs,Zhang:2005yz,Zhang:2006qu,Zhang:2007sh,Zhang:2009un,Zhang:2014nta,Zhang:2015rha,Cai:2015emx,Wang:2016lxa,Feng:2016djj,Guo:2015gpa,Guo:2018gyo,Poulin:2018cxd,Feng:2019jqa,DiValentino:2019jae,Yin:2023srb,Yao:2023qve,Wang:2024vmw,Giare:2024smz,CosmoVerseNetwork:2025alb,Kochappan:2024jyf,Li:2024qus,Liu:2025evk,Li:2025vuh,Li:2025owk,Song:2025bio,Xu:2026sbw,Zhang:2025lam}.

Early dark energy (EDE) is an important theoretical scenario proposed to alleviate the Hubble tension \citep{Poulin:2018cxd,Lin:2019qug,Smith:2019ihp,Ye:2020btb,Ye:2020oix,Braglia:2020bym,Agrawal:2019lmo,Poulin:2023lkg,Bella:2026zuk,Yin:2026gss,Wang:2026xws}. Its central idea is to introduce a transient energy component before recombination, increasing the early expansion rate and thereby reducing the sound horizon. A smaller sound horizon can accommodate a higher $H_0$ while preserving the precisely measured CMB acoustic angular scale \citep{Poulin:2018cxd,Knox:2019rjx}. In the canonical axion EDE model, the scalar field becomes dynamical near a critical redshift and subsequently dilutes rapidly, with its impact on the cosmic expansion history concentrated around the recombination epoch~\citep{Poulin:2018cxd}.

In recent years, the James Webb Space Telescope (JWST) has revealed a substantial population of luminous galaxies at high redshifts, whose ultraviolet luminosity function (UVLF) provide new observational information on early galaxy formation and the small-scale matter power spectrum \citep{Perez-Gonzalez:2023wta,Casey:2024uvlf,Donnan:2024uvlf,Harikane:2025uvlf}. Previous studies have shown that EDE can enhance the predicted abundance of early luminous galaxies and improve agreement with JWST observations \citep{Forconi:2023hsj,Shen:2024hpx,Huang:2024aog,Jiang:2024tll}. In our previous work, combining CMB observations, DESI DR2 baryon acoustic oscillation (BAO) measurements\footnote{The DESI results have hinted at dynamical dark energy, stimulating extensive investigations into early dark energy~\citep{Chaussidon:2025npr,Pang:2025lvh,Bella:2026zuk,Yin:2026gss,Wang:2024dka}, interacting dark energy~\citep{Li:2024qso,Giare:2024smz,Li:2025ula,Li:2025owk,Li:2026xaz,Pan:2025qwy,Wang:2025znm,Wang:2026wrk,Wu:2025vrl}, modified gravity~\citep{Yang:2024kdo,Ye:2024ywg,Wolf:2025jed,Du:2026cly}, cosmological neutrino masses~\citep{Du:2024pai,Du:2025xes,Jiang:2024viw,Elbers:2025vlz,Giare:2025ath,Du:2025iow,Li:2026mbh,Yang:2026qon,Yang:2026yaq}, and others~\citep{Li:2025dwz,Li:2026hwq,Liu:2026xfl,Giare:2026opg,Colgain:2026ryf}. For recent reviews, see, e.g., \citep{CosmoVerseNetwork:2025alb,Cai:2025mas,Li:2026asg,Anchordoqui:2026hys,Zhang:2026zbs,Giare:2026oti}.}~\citep{DESI:2025zgx}, and JWST high-redshift galaxy observations, we obtained $H_0=71.58\pm1.05~\rm km\,s^{-1}\,Mpc^{-1}$ in the axion EDE model, alleviating the tension with SH0ES to approximately $0.6\sigma$ \citep{Du:2026qtq}. This result raises a key physical question: why can small-scale structure information from high-redshift galaxies influence the inferred global expansion rate of the Universe and lead to such a high $H_0$ within the EDE framework?

In this work, we investigate this question from both observational and theoretical perspectives. We first account for the reference cosmologies adopted in the published JWST UVLF measurements and improve the likelihood construction \citep{Sabti:2021xvh}, obtaining $H_0=72.76\pm1.00~\rm km\,s^{-1}\,Mpc^{-1}$ in the axion EDE model, reducing the tension with SH0ES to the $0.1\sigma$ level. On the observational side, we find that the galaxy abundances at $z=11$ and $12.5$ lie significantly above the $\Lambda$CDM predictions, while excluding these measurements reduces $H_0$ by approximately $1$-$2~\rm km\,s^{-1}\,Mpc^{-1}$, demonstrating the crucial role of the highest-redshift galaxies in driving the higher $H_0$. On the theoretical side, these observations favor enhanced small-scale matter power, which can be accommodated in EDE by increasing the primordial scalar spectral index $n_{\rm s}$. Under the CMB constraints, the positive correlations among $n_{\rm s}$, $f_{\rm EDE}$, and $H_0$ consequently lead to a higher inferred $H_0$. By further introducing the running of the primordial scalar spectral index, $\alpha_{\rm s}$ \citep{Parashari:2023cui,AtacamaCosmologyTelescope:2025nti,Kobayashi:2026sji,Sabogal:2026qvy}, to allow greater scale-dependent freedom in the primordial power spectrum, we find that the joint observations favor a positive running at more than $4\sigma$, accompanied by a higher $H_0=73.75\pm1.04~\rm km\,s^{-1}\,Mpc^{-1}$. These results reveal the close connection among JWST high-redshift galaxy abundances, the shape of the primordial power spectrum, and the EDE model parameters, elucidating the physical mechanism by which high-redshift small-scale structure observations influence the inferred $H_0$ through the degeneracies between primordial power spectrum parameters and EDE parameters.

\section{Methodology and data}
\label{sec:methods}

\subsection{Axion EDE and the primordial spectrum}
\label{sec:model}

We adopt the canonical axion EDE model, characterized by the scalar field potential \citep{Poulin:2018cxd,Smith:2019ihp}
\begin{equation}
V(\phi)=m^2f^2\left[1-\cos\left(\frac{\phi}{f}\right)\right]^3,
\end{equation}
where $m$ and $f$ denote the mass and decay-constant scales of the scalar field, respectively. The field is initially frozen by Hubble friction and becomes dynamical near a critical redshift $z_{\rm c}$. We characterize the model by the fractional EDE energy density $f_{\rm EDE}$ at $z_{\rm c}$ and the initial field displacement $\theta_{\rm i}=\phi_{\rm i}/f$. The field subsequently oscillates with an average equation of state $\bar{w}_\phi=1/2$, so that its energy density dilutes as $\rho_\phi\propto a^{-9/2}$ during the oscillatory phase. Consequently, the contribution of EDE to the cosmic expansion history is concentrated around the recombination epoch.

For the primordial curvature power spectrum, we consider both the standard power-law form and an extension allowing the running of the scalar spectral index, expressed as
\begin{equation}
\ln\mathcal{P}_{\mathcal R}(k)
=\ln A_{\rm s}
+(n_{\rm s}-1)\ln\left(\frac{k}{k_*}\right)
+\frac{\alpha_{\rm s}}{2}\ln^2\left(\frac{k}{k_*}\right),
\label{eq:primordial}
\end{equation}
where $A_{\rm s}$ and $n_{\rm s}$ are the primordial amplitude and scalar spectral index at the pivot scale $k_*=0.05\,{\rm Mpc}^{-1}$, respectively, and $\alpha_{\rm s}\equiv {\rm d}n_{\rm s}/{\rm d}\ln k$ denotes the running of the scalar spectral index.

\begin{table*}[!t]
\centering
\renewcommand{\arraystretch}{1.25}
\setlength{\tabcolsep}{3.3pt}
\caption{Marginalized cosmological constraints for the baseline, data cuts, and running analyses using the CMB, DESI, and JWST data. The redshifts in the second block denote the \texttt{Donnan} bins removed from this combination~\citep{Donnan:2024uvlf}. For each data combination, the $Q_{\rm DMAP}$ tension is the square root of the increase in the best-fit \(\chi^2\) upon adding the SH0ES prior.}
\label{tab:results}
\begin{tabular}{lcccccc|c}
\hline\hline
Model / Data & $H_0\,[\rm km\,s^{-1}\,Mpc^{-1}]$ & $10^9 A_{\rm s}$ & $n_{\rm s}$ & $f_{\rm EDE}$ & $\alpha_{\rm s}$ & $r_{\rm d}\,[{\rm Mpc}]$ & $Q_{\rm DMAP}$ \\
\hline
\multicolumn{7}{l}{\textbf{Baseline}} & \\
$\Lambda$CDM & $68.20\pm 0.27$ & $2.181\pm 0.029$ & $0.9764\pm 0.0029$ & -- & -- & $147.56\pm 0.19$ & $4.5\sigma$ \\
EDE & $72.76\pm 1.00$ & $2.184^{+0.024}_{-0.027}$ & $0.9983^{+0.0059}_{-0.0074}$ & $0.143\pm 0.026$ & -- & $139.3\pm 1.6$ & $0.1\sigma$ \\
\hline
\multicolumn{7}{l}{\textbf{Data cuts}} & \\
EDE (no $z=12.5$) & $71.38\pm 1.13$ & $2.171\pm 0.026$ & $0.9902^{+0.0057}_{-0.0068}$ & $0.104^{+0.035}_{-0.031}$ & -- & $141.7\pm 2.0$ & $0.4\sigma$ \\
EDE (no $z=11,\,12.5$) & $70.50^{+0.98}_{-1.15}$ & $2.163^{+0.023}_{-0.027}$ & $0.9852^{+0.0054}_{-0.0061}$ & $0.075^{+0.031}_{-0.036}$ & -- & $143.2^{+2.0}_{-1.7}$ & $1.4\sigma$ \\
\hline
\multicolumn{7}{l}{\textbf{Running}} & \\
$\Lambda$CDM+$\alpha_{\rm s}$ & $68.07\pm 0.27$ & $2.152^{+0.027}_{-0.032}$ & $0.9762\pm 0.0030$ & -- & $0.0135\pm 0.0049$ & $147.66\pm 0.19$ & $4.7\sigma$ \\
EDE+$\alpha_{\rm s}$ (no JWST) & $69.80^{+0.85}_{-1.25}$ & $2.142^{+0.026}_{-0.030}$ & $0.9823^{+0.0054}_{-0.0078}$ & $< 0.120\ (95\%)$ & $0.0052^{+0.0056}_{-0.0069}$ & $144.4^{+2.3}_{-1.6}$ & $1.8\sigma$ \\
EDE+$\alpha_{\rm s}$ & $73.75\pm 1.04$ & $2.129\pm 0.027$ & $1.0142\pm 0.0070$ & $0.185\pm 0.028$ & $0.0317\pm 0.0074$ & $137.6\pm 1.7$ & $0.9\sigma$ \\
\hline
\end{tabular}
\end{table*}

\subsection{High-redshift luminosity functions}
\label{sec:jwst}

We use 39 JWST UVLF detections in the redshift range $7\leq z\leq12.5$ from \citet{Perez-Gonzalez:2023wta,Casey:2024uvlf,Donnan:2024uvlf,Harikane:2025uvlf}. Among these, 28 measurements belong to the \texttt{Donnan} sample \citep{Donnan:2024uvlf}, covering four redshift bins at $z=9$, 10, 11, and 12.5. To investigate the impact of different redshift bins on cosmological parameter inference, we construct two additional data combinations by excluding the 7 \texttt{Donnan} measurements at $z=12.5$, or the 14 measurements at $z=11$ and 12.5.

The theoretical UVLF can be expressed as
\begin{equation}
\Phi(M_{\rm UV},z)
=\frac{{\rm d}n}{{\rm d}M_{\rm UV}}
=\frac{{\rm d}n}{{\rm d}M_{\rm h}}
\frac{{\rm d}M_{\rm h}}{{\rm d}M_*}
\left|\frac{{\rm d}M_*}{{\rm d}M_{\rm UV}}\right|,
\label{eq:uvlf}
\end{equation}
where $M_{\rm h}$ and $M_*$ denote the halo mass and stellar mass, respectively. The first term, ${\rm d}n/{\rm d}M_{\rm h}$, is the halo mass function calculated using the Sheth-Mo-Tormen prescription \citep{Sheth:1999su}, which depends on the matter power spectrum. The second term is determined by a double power-law stellar-to-halo mass relation \citep{Stefanon:2021shmr,CEERSTeam:2023qgy,Scholtz:2023}, while the third term is calculated using the redshift-dependent stellar mass-UV magnitude relation of \citet{Stefanon:2021shmr}, including dust attenuation corrections based on the UV continuum slope \citep{Meurer:1999jj,Cullen:2023uv}.

In this work, we further account for the reference cosmologies adopted by different observational samples. Since the inferred absolute UV magnitude $M_{\rm UV}$ and comoving number density $\Phi$ depend on the cosmological model assumed in constructing the UVLF, we apply the cosmology-dependent conversions described by \citet{Sabti:2021xvh}. The absolute UV magnitude is converted according to
\begin{equation}
M_{\rm UV}=M_{\rm UV}^{\rm ref}
-5\log_{10}\left[
\frac{D_{\rm L}(z)}{D_{\rm L}^{\rm ref}(z)}
\right],
\label{eq:mag_conversion}
\end{equation}
where $D_{\rm L}$ is the luminosity distance in the cosmological model being tested, and the superscript $\mathrm{ref}$ denotes the reference cosmology adopted by the observational sample. Correspondingly, the theoretical UVLF is converted using the comoving volume ratio,
\begin{equation}
\Phi_{\rm th}^{\rm ref}(M_{\rm UV}^{\rm ref},z)
=\frac{V(z_-,z_+)}{V^{\rm ref}(z_-,z_+)}
\Phi_{\rm th}(M_{\rm UV},z),
\label{eq:ref}
\end{equation}
where $V(z_-,z_+)$ is the comoving volume within the redshift interval $[z_-,z_+]$. Through these conversions, we consistently transform the theoretical predictions of different cosmological models into the reference cosmology adopted by each observational sample, allowing a consistent comparison with the JWST UVLF data.

\begin{figure*}[!t]
\centering
\includegraphics[width=1.0\textwidth]{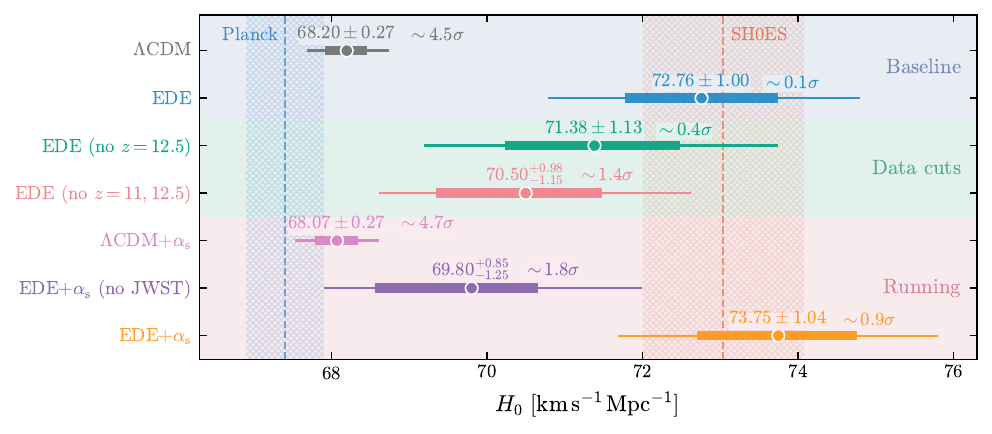}
\caption{Overview of the marginalized $H_0$ constraints from the baseline, high-redshift \texttt{Donnan} UVLF measurements data-cut, and primordial spectral running analyses. The thick and thin error bars represent the 68\% and 95\% credible intervals, respectively. The $N\sigma$ values give the $Q_{\rm DMAP}$ tension with SH0ES.}
\label{fig1}
\end{figure*}

\subsection{Data and bayesian inference}

For the CMB observations, we combine the Planck 2018 low-$\ell$ temperature and polarization data with the high-$\ell$ power spectra from Planck, ACT DR6, and SPT-3G D1 \citep{Planck:2019nip,AtacamaCosmologyTelescope:2025blo,SPT-3G:2025bzu}. To avoid double counting between experiments, we adopt the ACT-compatible $\ell$ cuts of $\ell_{\max}=1000$ for the Planck TT spectrum and $\ell_{\max}=600$ for the TE and EE spectra. We also include the joint CMB lensing power spectrum reconstructed from Planck PR4, ACT DR6, and SPT-3G \citep{ACT:2025qjh}. For the BAO observations, we adopt the DESI DR2 measurements~\citep{DESI:2025zgx}.

We compute the background and linear perturbation evolution using a modified version of \texttt{CAMB} \citep{Lewis:1999bs}. We perform Markov chain Monte Carlo sampling using \texttt{Cobaya} \citep{Torrado:2020dgo}, analyze the resulting chains with \texttt{GetDist} \citep{Lewis:2019xzd}, and assess convergence using the Gelman--Rubin criterion $R-1<0.02$ \citep{Gelman:1992zz}.

\section{Results and discussion}
\label{sec:results}

Figure~\ref{fig1} and Table~\ref{tab:results} summarize our main results. We quantify the residual tension with SH0ES using $Q_{\rm DMAP}$ \citep{Raveri:2018wln,Schoneberg:2021qvd}, the square root of the increase in the minimum $\chi^2$ after adding the SH0ES prior. With the revised JWST UVLF analysis, the baseline CMB+DESI+JWST data yield $H_0=72.76\pm1.00~\rm km\,s^{-1}\,Mpc^{-1}$ in the axion EDE model, in excellent agreement with SH0ES at the $0.1\sigma$ level. In the following, we investigate the physical origin of this higher $H_0$.

\subsection{The role of the highest-redshift JWST bins}
\label{sec:cuts}

\begin{figure}[!t]
\centering
\includegraphics[width=0.47\textwidth]{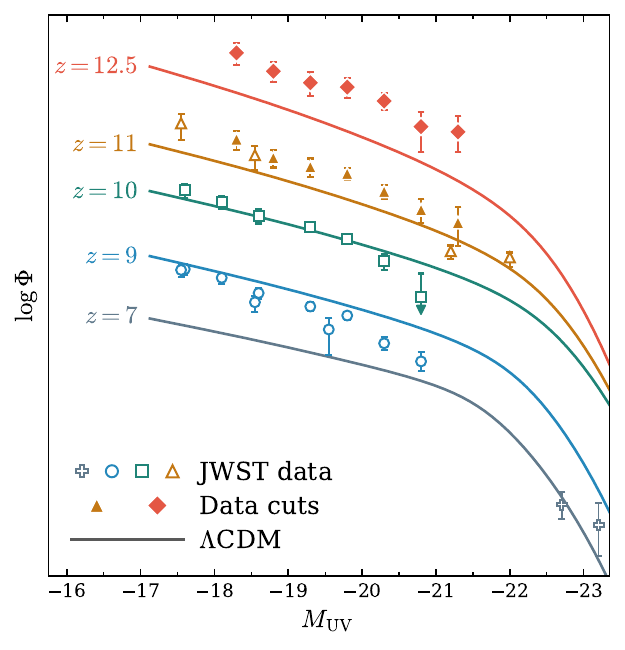}
\caption{JWST UVLF observations at different redshifts compared with the theoretical predictions of the $\Lambda$CDM model from CMB+DESI+JWST. The filled symbols highlight the \texttt{Donnan} measurements at $z=11$ and $12.5$ considered in the data-cut analyses, while the open symbols represent the remaining JWST measurements. Note that redshift-dependent offsets are applied for visual clarity, preserving the relative positions of the observations and theoretical predictions.}
\label{fig2}
\end{figure}

\begin{figure*}[!t]
\centering
\includegraphics[width=\textwidth]{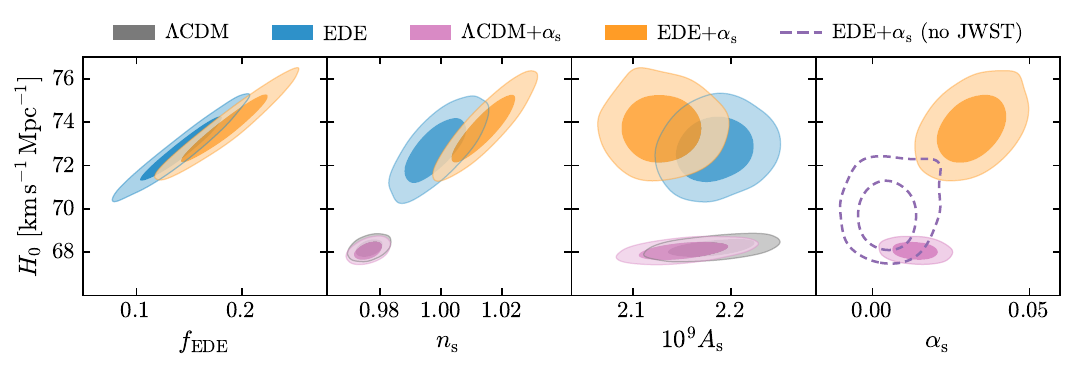}
\caption{Two-dimensional marginalized contours of $H_0$ with $f_{\rm EDE}$, $n_{\rm s}$, $A_{\rm s}$, and $\alpha_{\rm s}$ from the CMB, DESI, and JWST data.}
\label{fig3}
\end{figure*}

\begin{figure*}[!t]
\centering
\includegraphics[width=0.33\textwidth]{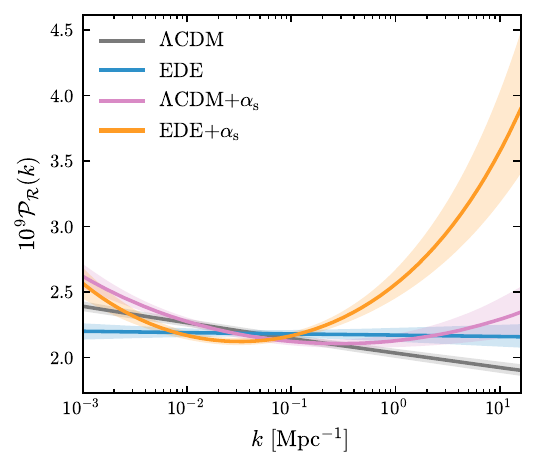}
\includegraphics[width=0.33\textwidth]{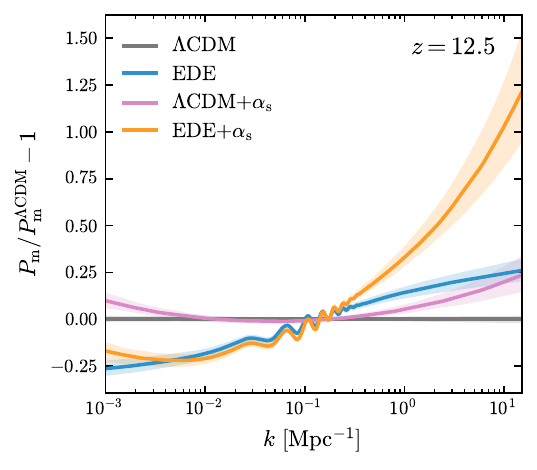}
\includegraphics[width=0.33\textwidth]{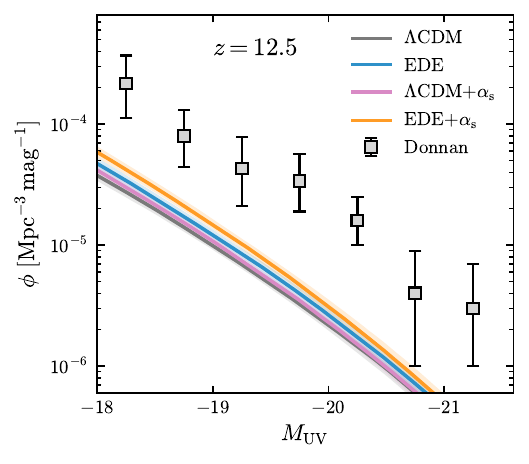}
\caption{Primordial and matter power spectra and UVLF predictions for the baseline and running models. \textit{Left panel:} Primordial power spectra with posterior medians and 68\% credible intervals. \textit{Middle panel:} Fractional differences in the matter power spectra relative to $\Lambda$CDM with 68\% credible intervals. \textit{Right panel:} UVLF predictions at $z=12.5$ with 68\% credible intervals, compared with the JWST observations from \citet{Donnan:2024uvlf}.}
\label{fig4}
\end{figure*}

To identify which JWST observations play a key role in the inference of $H_0$, we first compare the observed UVLF at different redshifts with the theoretical predictions of the $\Lambda$CDM model. As shown in Figure~\ref{fig2}, the observations at $z=7$, 9, and 10 are generally consistent with the $\Lambda$CDM predictions, whereas the galaxy abundances at the highest redshifts, $z=11$ and 12.5, lie systematically above the theoretical predictions. These results indicate that the highest-redshift galaxy observations favor substantially stronger matter clustering. To examine the impact of these high-redshift measurements, we further perform data-cut tests by excluding the corresponding \texttt{Donnan} measurements.

As shown in Figure~\ref{fig1} and Table~\ref{tab:results}, excluding only the $z=12.5$ measurements reduces $H_0$ from $72.76$ to $71.38~\rm km\,s^{-1}\,Mpc^{-1}$, while removing both the $z=11$ and $12.5$ bins further lowers it to $70.50~\rm km\,s^{-1}\,Mpc^{-1}$. Correspondingly, the Hubble tension with SH0ES increases from $0.1\sigma$ to $0.4\sigma$ and $1.4\sigma$, respectively. Meanwhile, $f_{\rm EDE}$ decreases from $0.143\pm0.026$ to $0.075^{+0.031}_{-0.036}$, and $n_{\rm s}$ decreases from approximately $0.998$ to $0.985$. These results demonstrate that the highest-redshift JWST observations play a crucial role in driving the larger $H_0$.

\subsection{Small-scale matter power and the EDE parameter degeneracy}
\label{sec:mechanism}

The above results demonstrate a close connection between the enhanced abundance of high-redshift galaxies and the higher $H_0$ inferred in the EDE model. Luminous galaxies at high redshifts are generally hosted by relatively rare, massive dark matter halos, whose abundance is highly sensitive to the amplitude of matter density fluctuations. A higher galaxy abundance can generally result from stronger small-scale matter power. For a power-law primordial spectrum, increasing the scalar spectral index $n_{\rm s}$ enhances the primordial power at wavenumbers above the pivot scale, leading to stronger matter clustering on the corresponding scales. Consequently, the enhanced galaxy abundances observed by JWST at $z=11$ and $12.5$ favor a larger $n_{\rm s}$ in the joint analysis.

On the other hand, $n_{\rm s}$ is tightly constrained by the CMB observations. In EDE models, a higher $n_{\rm s}$ is generally associated with a larger $H_0$, and this positive correlation can be understood from the CMB temperature anisotropy power spectrum. For a power-law primordial spectrum, the Sachs-Wolfe contribution approximately satisfies \citep{Ye:2020btb}
\begin{equation}
(C_\ell)_{\rm SW}\propto A_{\rm s}(D_A^*k_{*})^{1-n_{\rm s}}f(n_{\rm s}),
\label{eq:sw_ede}
\end{equation}
where $D_A^*$ is the comoving distance to the last-scattering surface, and $f(n_{\rm s})$ is a monotonically increasing function of $n_{\rm s}$. In EDE models, a higher $H_0$ corresponds to a smaller $D_A^*$, which reduces the factor $(D_A^*k_*)^{1-n_{\rm s}}$ for $n_{\rm s}<1$. To maintain consistency with the observed CMB power spectrum, a larger $n_{\rm s}$ is therefore favored.

As shown in Figure~\ref{fig3}, our analysis clearly exhibits this positive correlation, together with a strong positive correlation between $H_0$ and $f_{\rm EDE}$. Therefore, when the JWST high-redshift galaxy observations require enhanced small-scale matter power, the joint fit can increase $n_{\rm s}$ to enhance the primordial power on these scales, while moving along the parameter degeneracy allowed by the CMB toward larger $f_{\rm EDE}$ and $H_0$. This mechanism explains why small-scale galaxy observations can substantially influence the inference of the global expansion rate of the Universe.

\subsection{The impact of running of the scalar spectral index}
\label{sec:running}

To further test whether the enhancement of the small-scale primordial power spectrum is an important mechanism driving the increase in $H_0$, we introduce the running of the scalar spectral index, $\alpha_{\rm s}$, allowing greater scale-dependent freedom in the primordial power spectrum. As shown in Table~\ref{tab:results}, the EDE+$\alpha_{\rm s}$ model yields $H_0=73.75\pm1.04~\rm km\,s^{-1}\,Mpc^{-1}$ from the CMB+DESI+JWST data. Meanwhile, the joint data favor a significantly positive running, $\alpha_{\rm s}=0.0317\pm0.0074$, corresponding to a deviation from zero of approximately $4.3\sigma$. The scalar spectral index further increases to $n_{\rm s}=1.0142\pm0.0070$, exceeding the scale-invariant value by approximately $2.0\sigma$, while $f_{\rm EDE}$ increases to $0.185\pm0.028$. In contrast, using only the CMB+DESI data gives a running consistent with zero and $H_0\simeq69.80~\rm km\,s^{-1}\,Mpc^{-1}$.

Figure~\ref{fig4} further illustrates the impact of primordial spectral running on small-scale power and galaxy abundances. As shown in the left panel, a positive $\alpha_{\rm s}$ in the EDE+$\alpha_{\rm s}$ model significantly enhances the primordial power on small scales, with the enhancement increasing rapidly toward larger $k$. At $k=10\,\mathrm{Mpc}^{-1}$, the primordial power increases by approximately 70\% relative to the baseline EDE model. The middle panel further shows that at $k=10\,\mathrm{Mpc}^{-1}$, the matter power in the EDE+$\alpha_{\rm s}$ model is enhanced by more than 100\% relative to $\Lambda$CDM, substantially exceeding the enhancement in the baseline EDE model. Correspondingly, the right panel shows that the enhanced small-scale power increases the predicted galaxy abundance at $z=12.5$, bringing it closer to the JWST observations. These results indicate that increasing $n_{\rm s}$ alone within the baseline EDE framework is insufficient to fully account for the small-scale power favored by JWST. Introducing additional scale-dependent freedom allows the joint data to favor a larger $n_{\rm s}$ and a positive $\alpha_{\rm s}$, further enhancing small-scale power and driving $H_0$ to higher values.

\section{Conclusion}
\label{sec:conclusion}

In this work, we investigate the physical mechanism by which JWST high-redshift galaxy observations favor a higher $H_0$ within the axion EDE framework from both observational and theoretical perspectives.

By accounting for the reference cosmologies of the JWST UVLF measurements, we obtain $H_0=72.76\pm1.00~\mathrm{km\,s^{-1}\,Mpc^{-1}}$ in the axion EDE model, reducing the Hubble tension with SH0ES to the $0.1\sigma$ level. The observed galaxy abundances at $z=11$ and $12.5$ lie significantly above the $\Lambda$CDM predictions, while excluding these measurements reduces $H_0$ by approximately $2.26~\mathrm{km\,s^{-1}\,Mpc^{-1}}$, increasing the tension with SH0ES to as much as $1.4\sigma$. These results demonstrate their crucial role in driving the higher $H_0$. The observations favor stronger small-scale matter power, which drives the joint fit toward a larger primordial scalar spectral index $n_{\rm s}$. Under the CMB constraints, the positive correlations among $n_{\rm s}$, $f_{\rm EDE}$, and $H_0$ translate this requirement into a higher inferred $H_0$.

By further introducing the running of the primordial scalar spectral index, $\alpha_{\rm s}$, to allow greater scale-dependent freedom in the primordial power spectrum, we find that the joint observations favor a positive running at more than $4\sigma$, accompanied by further increases in $n_{\rm s}$ and $f_{\rm EDE}$, raising $H_0$ to $73.75\pm1.04~\mathrm{km\,s^{-1}\,Mpc^{-1}}$. This result provides further support for the important role of enhanced small-scale primordial power in driving the higher $H_0$ within the EDE framework.

Overall, our findings reveal the close connection among JWST high-redshift galaxy abundances, the shape of the primordial power spectrum, and the Hubble constant, elucidating the physical mechanism by which high-redshift small-scale structure observations influence the inferred $H_0$. These results provide a systematic physical understanding of the connection between JWST observations and the Hubble tension.

\section*{Acknowledgments}
We thank Lu Yin, Sheng-Han Zhou, Yun-He Li and Yi-Min Zhang for their helpful discussions. This work was supported by the National Natural Science Foundation of China (Grants Nos. 12533001, 12575049, and 12473001), the National SKA Program of China (Grants Nos. 2022SKA0110200 and 2022SKA0110203), the China Manned Space Program (Grant No. CMS-CSST-2025-A02), and the 111 Project (Grant No. B16009). This work is based [in part] on observations made with the NASA/ESA/CSA James Webb Space Telescope (JWST) and Hubble Space Telescope (HST). The data were obtained from the Mikulski Archive for Space Telescopes (MAST) at the Space Telescope Science Institute, which is operated by the Association of Universities for Research in Astronomy, Inc., under NASA contract NAS 5-03127 for JWST and NAS 5-26555 for HST. These observations are associated with programs 3293, 1837, 1180, 1210, and 2079, and accessed via \dataset[10.17909/je9x-d314]{https://doi.org/10.17909/je9x-d314} and \dataset[10.17909/gbkc-2b17]{https://doi.org/10.17909/gbkc-2b17}. We also acknowledge data obtained from the ESO Science Archive Facility with \dataset[10.18727/archive/52]{https://doi.org/10.18727/archive/52}. OpenAI's ChatGPT and Codex (GPT-6 Astra) were used to improve the language of the manuscript, cross-check algebraic calculations and derivations, and assist with code modification and numerical checks. The authors take full responsibility for the scientific content and conclusions of this work.

\bibliographystyle{aasjournal}
\bibliography{main}
\end{document}